\documentclass[10pt,twocolumn,aps,prl,superscriptaddress,reprint]{revtex4-1}

\usepackage[latin9]{inputenc}
\usepackage{bm}
\usepackage{amsmath}
\usepackage{graphicx}
\usepackage[unicode=true,
 bookmarks=false,
 breaklinks=false,pdfborder={0 0 1},colorlinks=false]
 {hyperref}
\usepackage{breakurl}

\usepackage{float}
\usepackage{threeparttable} 
\usepackage{multirow}

\makeatletter

\usepackage{xcolor}
\usepackage{bm}
\usepackage{url}
\usepackage{ulem}

\makeatother

\newcommand{\be}{\begin{equation}}
\newcommand{\ee}{\end{equation}}

\newcommand{\bk}{{\bm{k}}}

\newcommand{\bq}{{\bm{q}}}

\renewcommand{\thefootnote}{\roman{footnote}}

\begin{document}

\title{Emergent superconductivity in two-dimensional NiTe$_2$ crystals}

\author{Feipeng Zheng}
\email{fpzheng_phy@jnu.edu.cn}
\affiliation{Siyuan Laboratory, Guangzhou Key Laboratory of Vacuum Coating Technologies and New Energy Materials, Department of Physics, Jinan University, Guangzhou 510632, China}

\author{Xi-Bo Li} 
\affiliation{Siyuan Laboratory, Guangzhou Key Laboratory of Vacuum Coating Technologies and New Energy Materials, Department of Physics, Jinan University, Guangzhou 510632, China}

\author{Yiping Lin}
\affiliation{Siyuan Laboratory, Guangzhou Key Laboratory of Vacuum Coating Technologies and New Energy Materials, Department of Physics, Jinan University, Guangzhou 510632, China}

\author{Lingxiao Xiong}
\affiliation{Siyuan Laboratory, Guangzhou Key Laboratory of Vacuum Coating Technologies and New Energy Materials, Department of Physics, Jinan University, Guangzhou 510632, China}

\author{Xiaobo Chen} 
\affiliation{Siyuan Laboratory, Guangzhou Key Laboratory of Vacuum Coating Technologies and New Energy Materials, Department of Physics, Jinan University, Guangzhou 510632, China}

\author{Ji Feng}
\email{jfeng11@pku.edu.cn}
\affiliation{International Center for Quantum Materials, School of Physics, Peking
University, Beijing 100871, China}
\affiliation{Collaborative Innovation Center of Quantum Matter, Beijing 100871,
China}

\date{\today}
\begin{abstract}
Despite growing interest in them, highly crystalline two-dimensional superconductors derived from exfoliated layered materials are few. Employing the anisotropic Migdal-Eliashberg formalism based on {\it ab initio} calculations, we find monolayer NiTe$_{2}$ to be an intrinsic superconductor with a  $T_{\text c}\sim$5.7~K, although the bulk crystal is not known to superconduct. Remarkably, bilayer NiTe$_{2}$ intercalated with lithium is found to display two-gap superconductivity with a critical temperature $T_{\text{c}}\sim 11.3$~K and superconducting gap of $\sim$3.1~meV, arising from a synergy of electronic and phononic effects. The comparatively high $T_\text{c}$, substrate independence and proximity tunability will make these superconductors ideal platforms for exploring intriguing correlation effects and quantum criticality associated two-dimensional superconductivity.
\end{abstract}
\maketitle

Highly crystalline two-dimensional superconductors derived from exfoliated layered materials represent a  unique class of two-dimensional superconductivity without the need for an indispensable substrate.~\citep{Cao15, Xi15, Jiang14, Navarro-Moratalla16,Ugeda15,Cao18a,Cao18b,Yu19} In spite of the fascinating physics associated with them, the discovery of these exfoliated two-dimensional superconductors has been few and far between.~\citep{Jiang14,Cao15,Xi15,Navarro-Moratalla16,Yu19} In particular, monolayer transition metal dichalcogenides,  NbSe$_2$ and  TaS$_2$, thinned down to the monolayer limit display coexisting superconductivity and charge-density wave driven by electron-phonon coupling.~\citep{Ugeda15,Navarro-Moratalla16,Xi16,Zheng18,Zheng19}
These two-dimensional crystals can be transferred from one substrate to another, and the superconductivity persists subject to perhaps but mild tuning by the substrate in proximity. Thus, this class of highly crystalline superconductors harbor truly two-dimensional and freestanding superconductivity, offering an ideal platform for exploring the interplay of novel superconductivity, quantum criticality and electron correlation effects, as well as novel superconducting device systems.\citep{Jiang14, Cao15, Xi15, Ugeda15, Navarro-Moratalla16, Xi16, Wang17, Cao18a, Cao18b, Zheng18, Zheng19} 

Indeed, the aforementioned NbSe$_2$ and  TaS$_2$ in its bulk crystalline form are archetypical layered transition metal dichalcogenides, both of which undergoes a phase transition to a commensurate charge-density wave phase  and then another into an anisotropic $s$-wave superconductors at the superconducting transition temperature $T_{\text{c}}$. And they are about the only known examples of intrinsic two-dimensional superconductors exfoliated from layered transition metal dichalcogenides. Clearly, the availability of superconducting phase in the true two-dimensional limit will help shed light on the decades-old puzzle regarding the coexistence and interplay of these competing orders,\citep{Ugeda15}  in addition to the exploration of novel physics in two-dimensional superconducting phases. There also appears to be an \textit{imminent} spin-density wave when NbSe$_2$ is thinned to monolayer limit in addition to the coexistent superconductivity and charge-density wave,~\citep{Zheng18,Zheng19} which may be suggesting a highly curious possibility that two-dimensional crystalline superconductors may have a phase diagram  resembling those in the high-$T_{\text c}$ class. The lack of such materials hinders the exploration of such phase diagrams. Therefore, it is highly desirable to find more examples of two-dimensional crystalline superconductor, to bring forth such an intriguing possibility while examining all the aforementioned fascinating physics.

This paper reports a computational investigation of the superconductivity in two-dimensional crystals of a transition metal dichalcogenides NiTe$_2$, where the superconducting temperature is calculated based on the anisotropic Midgal-Eliashberg formalism. Recently, NiTe$_{2}$ two-dimensional crystals have been successfully prepared with an accurately  number of layers down to monolayer limit~\cite{Zhao18}. Although the bulk NiTe$_2$ is not known to superconduct, we find that a two-gap superconductivity emerges in the the monolayer limit, with a T$_{\text c} \sim 5.7$~K. We show that Li-intercalated bilayer NiTe$_{2}$ tends to form the structure where all the intercalated sites are occupied by Li. Surprisingly, the superconductivity disappears in a bilayer geometry,  but when the bilayer NiTe$_2$ is intercalated with alkali metals a two-gap  superconductor re-emerges  with $T_{\text{c}}$ as high as 11.3 K and superconducting gap up to 3.1~meV. The superconducting mechanism and effects of Li intercalation are analyzed in detail. The $T_{\text{c}}$ can be further enhanced by in-plain compressive strain and electron doping, suggesting  the substrate with smaller lattice constant and higher work function can help to promote $T_{\text{c}}$ of monolayer NiTe$_{2}$.

\begin{figure}[H]
	\centering
	\includegraphics[width=75 mm]{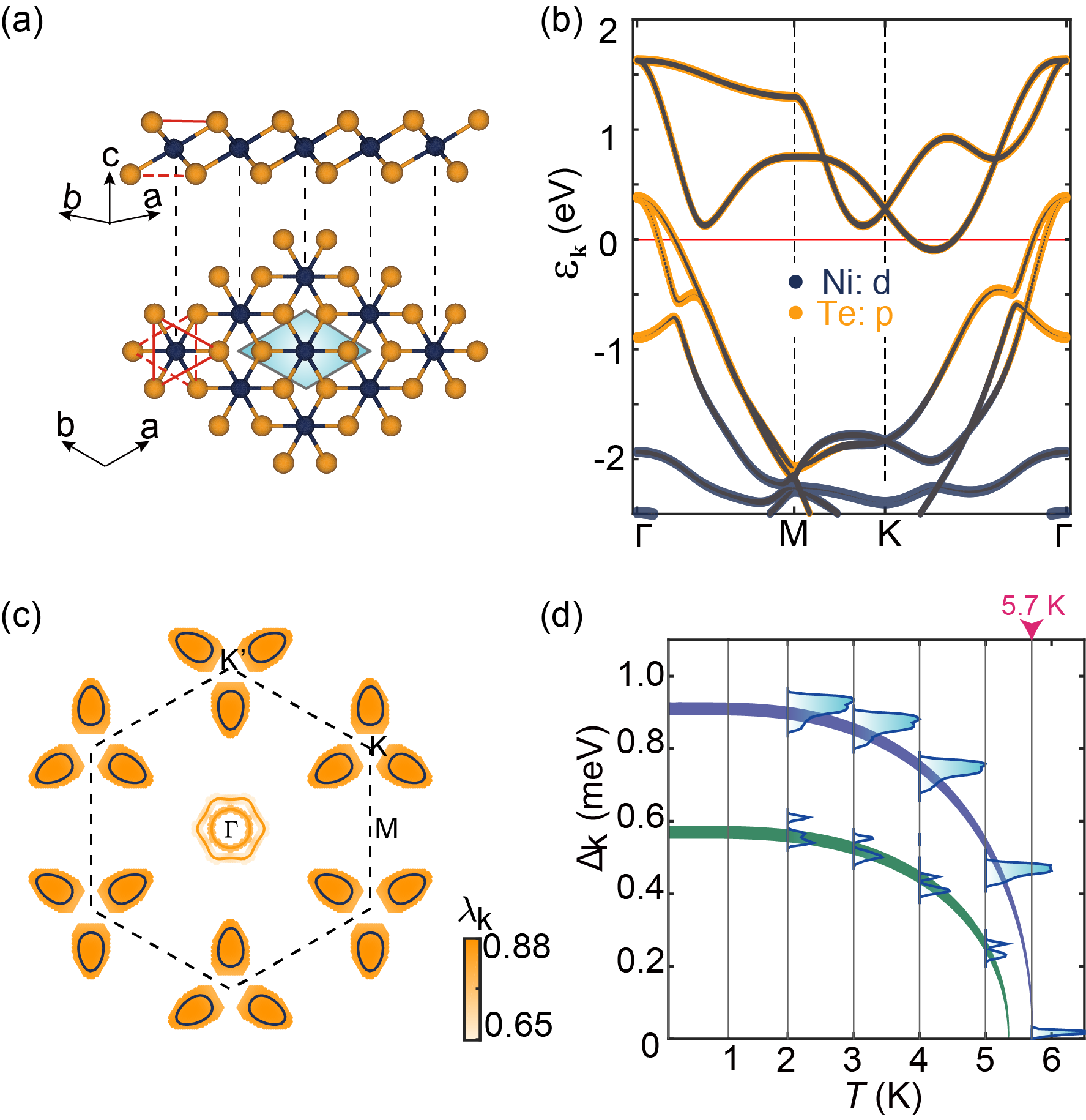} \\	
		\caption{ (a) Side (top) and top (bottom) views  of crystal structure of monolayer NiTe$_{2}$. The filled rhombus corresponds to a primitive unit cell.  (b) Band structure with projections onto constituent atoms, and (c) Fermi surface (red lines) of the monolayer superimposed with the distribution of $\lambda_{\bk}$ around Fermi surface. (d) Histograms of  $\Delta_{\bk}(\omega)$, evaluated at $T$'s from 2 to 6~K. The blue and green curves are BCS fits. }
	\label{fig1}
\end{figure}

In a single layer NiTe$_2$, each Ni atom is situated in the center of an octahedral cage formed by Te as shown in Fig.~\ref{fig1}(a). Te atoms occupy vertices of trigonally squashed octahedra, which then share edges to inflate into a two-dimensional sheet, with the 1$T$-type structure. Multilayer NiTe$_2$ and bulk crystals are composed of AA stacked monolayers. It is noted that the interlayer gallery of NiTe$_2$ has a spacing of 2.63 \AA, which is considerably smaller than other transition metal dichalcogenides (e.g. 3.08~\AA~ for TiTe$_{2}$, 2.89~\AA~ for WTe$_{2}$, 2.90~\AA~for NbSe$_{2}$), indicating relatively strong interlayer coupling. Density-functional theory and density-functional perturbation theory  calculations are performed to study the crystal, electronic, phonon structures and electron-phonon coupling  of few-layer NiTe$_{2}$~\citep{Baroni87,Giannozzi09,Hamann79,Perdew96}.  The Kohn-Sham valence states on a $18\times 18$ $\bk$-grid are expanded as planewaves below 100 Rydberg, whereupon the geometric and electronic structures are evaluated.

The optimized structure of bulk NiTe$_2$ using different types of exchange-correlation functionals available are carefully examined. It is found that both the standard generalized-gradient PBE functional,\citep{Perdew96} and the PBE functional incorporating long-range van der Waals interaction~\citep{Hamada14} predict lattice constants sufficiently close to the experiments.\citep{Bensch96,Furuseth65,SM} The optimized $a$ value for monolayer NiTe$_2$ is 3.772 \AA, which is smaller than bulk, but becomes very close to the bulk value for multilayer NiTe$_2$. 
We begin by highlighting the main features of the electronic and phonon band structures in monolayer NiTe$_{2}$. Figs.~\ref{fig1}(b) and~\ref{fig1}(c)  display the  projected bandstructure and Fermi surface. Three bands across Fermi level, $\varepsilon_{\text{F}}$, which makes the monolayer a compensated metal, as shown in Fig.~\ref{fig1}(b). There are two hole pockets at $\Gamma$, ascribable mainly to the $p$-orbitals on Te. At the corners of the Brillouin zone, K and K$'$, are clover-shaped electron pockets with three separate petals each,~[Fig.~\ref{fig1}(c)] which are admixture of $d$-orbitals of Ni and $p$-orbitals of Te [Fig.~\ref{fig1}(b)]. The phonon spectrum shows~\citep{SM} no soft modes, indicating the stability of the optimized crystal structure, especially against charge-density waves. The projected phonon density of states~\citep{SM} reveals that the optical phonons are segregated into two rather narrow sectors with a 12 meV gap between them, featuring respectively Te-dominated modes at 10 -- 16 meV and Ni dominated modes 28 -- 30 meV.

In order to assess the superconductivity of monolayer NiTe$_{2}$, we employ the imaginary-time anisotropic Migdal-Eliashberg formalism~\citep{Migdal58, Eliashberg60} with subsequent analytic continuation to the real axis  using Pad$\acute{\textrm{e}}$ functions, which provides a quantitatively adequate description of electron-phonon coupling driven superconductivity in layered crystals,~\citep{Choi02,Margine13} with which the superconducting gap $\Delta(\bm k, \omega)$ is obtained by solving~\citep{Ponce2016,Margine13} the gap equation at finite temperatures.  Only the Kohn-Sham states within 100 meV of $\varepsilon_{\text{F}}$ are included, and the Matsubara frequencies are cut off at 0.32 eV. Electron-phonon coupling matrix elements are first computed on the above $\bk$- and $\bq$-grids, then interpolated~\cite{mostofi2008wannier90} to grids of $120\times120$.  The Coulomb pseudopotential $\mu^*$ is estimated to be 0.17 according to $\mu^{*}\approx [0.26N(0)/(1+ N(0) )]$,~\citep{Garland72} where $N$(0) is the electronic density of states at  $\varepsilon_{\text{F}}$.  

The Migdal-Eliashberg equations are then solved at finite temperatures for the superconducting gap $\Delta(\bm k, \omega)$~\citep{Giustino17} whereupon the $T_{\text c}$ is determined. The $\bk$-resolved superconducting gap on the Fermi surface, $\Delta_{\bk}(T)$,  for temperatures between 2 and 6 $K$ and $\omega =0$, are displayed in Fig. \ref{fig1}(d). It is seen that the gap vanishes at around 5.7 K, already higher than the liquid helium temperature. Below the transition temperature, we see that the superconductivity shows a patent two-gap feature, a point to be returned to shortly. It is  worth remarking that the $T_{\text c}$ obtained with the McMillan-Allen-Dynes (MAD) approach,~\citep{McMillan68,Allen75,Giustino17} which assumes an isotropic spectrum, is around 2.3 K for the monolayer. Solving the full Midgal-Eliashberg gap equation is evidently more reliable given the reduced dimensionality and anisotropic Fermi surface.\citep{Choi02,Margine13,Sanna12} It is also noted that the superconducting transition $T_{\text{c}}$ computed using  MAD and Migdal-Eliashberg approaches  for bulk NiTe$_2$ are both very close to zero temperature, consistent with the fact that bulk NiTe$_2$ has not been reported to superconduct. The in-plane compressive strain and electron doping can further enchance $T_{\text{c}}$ of the monolayer.\citep{SM} 

Interestingly, our calculations also show that the estimated $T_{\text{c}}$'s using both the MAD and Midgal-Eliashberg of bilayer NiTe$_{2}$ are below 1~K.\citep{SM} This and the absence of superconductivity in bulk  NiTe$_{2}$  possibly indicate that the reduced interlayer spacing and concomitant strong interlayer coupling, as mentioned earlier, in this layered material is detrimental to superconductivity. This observation naturally begets the question whether the superconductivity seen in monolayer can be reconstituted if we can weaken the interlayer coupling by, for example, augmenting interlayer spacing. Indeed, the interlayer space of layered structures is a natural gallery for a wide range of chemical species, through a chemical process called intercalation whereupon the interlayer space gets expanded by the intercalant.\citep{Mueller-Warmuth12} Here, we choose a family of simplest atomic intercalant, alkali metals, which can be inserted into few-layer NiTe$_2$ via ionic liquid gating. We shall focus mostly on lithium insertion in the following discussion, as it produces most dramatic boost to the superconducting $T_{\text c}$.\citep{SM} 

\begin{figure}[t]
\centering
\includegraphics[width=76 mm]{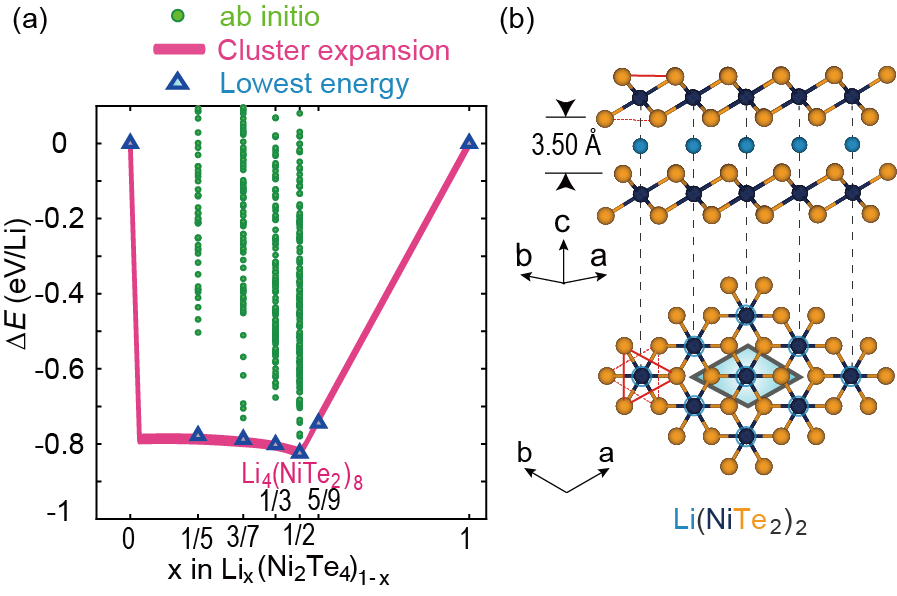} 
\caption{ (a) Computed formation energies for Li$_{x}$(Ni$_{2}$Te$_{4}$)$_{1-x}$. The green dots are the energies of structures sampled by ab initio method (DFT), with blue triangles indicating the lowest-energy structures. The magenta outline the convex hull from cluster expansion. (b) Side view (upper) and top view (lower) of the crystal structure of LiNi$_{2}$Te$_{4}$. The light-blue circles indicate intercalation sites. }
\label{fig2}
\end{figure} 
The energetically most favored lithium intercalated NiTe$_2$ bilayer has a stoichiometry, LiNi$_2$Te$_4$, of which the structure is displayed in Fig. \ref{fig2}(b). Li atoms are located precisely in between two eclipsed Ni from the two monolayers, forming a two-dimensional hexagonal lattice. The stoichiometry and structure are arrived at from the zero-temperature formation energies computed for a total of 36200 candidate lithium intercalation structures at multiple stoichiometries, combining {\it ab initio} particle swam optimization~\cite{Wang10} and random sampling of structures using a Hamiltonian based on cluster expansion.\citep{SM} The resultant convex hull of formation energy lithium-intercalated bilayer NiTe$_2$ shown in Fig. \ref{fig2}(a), pointing to the aforementioned stoichiometry and structure as thermodynamically stable against decomposition to any other stoichiometry. The interlayer gallery in LiNi$_2$Te$_4$ is substantially expanded to 3.50 \AA, as compared to 2.67 \AA~in the pristine bilayer. 

\begin{figure}[b]
\centering
\includegraphics[width=76 mm]{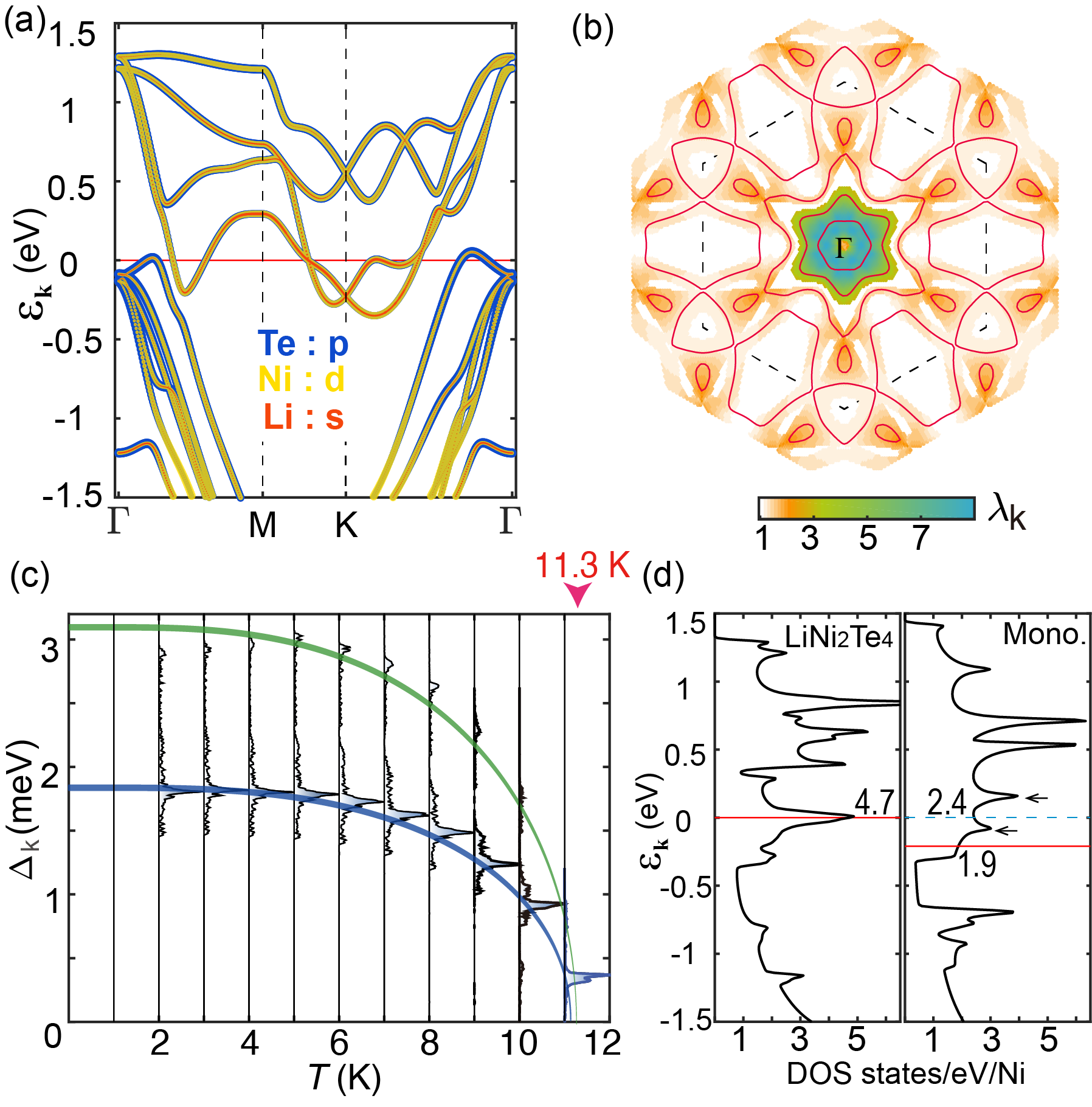} \\	
\caption{ (a) The band structure with projections onto constituent atoms and (b) Fermi surfaces (red lines) for LiNi$_{2}$Te$_{4}$,  superimposed with $\lambda_{\bk}$ around Fermi surface. (c) Histograms  of  $\Delta_{\bk}$ of Ni$_{2}$LiTe$_{4}$ at various temperatures. Blue and green curves are BCS fits of the two gaps. (d) Electronic density of states of LiNi$_{2}$Te$_{4}$ (left) and monolayer NiTe$_{2}$ (right). The solid red lines indicate Fermi levels.}
\label{fig3}
\end{figure}

The low-energy electronic excitations of LiNi$_2$Te$_4$ bears essential similarity with the pristine monolayer NiTe$_2$, but with a few remarkable departures. As shown in  Fig.~\ref{fig3}(a), the tortuous $t_{2g}$-$e_g$ gap is by and large preserved, although there is a substantial electron doping into the NiTe$_2$'s $e_g$ bands, the bottom of which also hybridize with lithium orbitals. Due to electron transfer from Li, large hole pockets entering M and $\Gamma$ points emerge, showing simultaneous Ni, Te and Li hybridization. The $p$-$d$ hybridized electron pockets at K and K$'$ are also slightly enlarged and three petals become fused at a pistil, as shown in Fig. \ref{fig3}(b).  The $t_{2g}$ bands show splittings owing to interlayer coupling mediated by the intercalated lithium, whereas The hole pockets at $\Gamma$ are still dominated by $p$-orbitals of Te.  

Remarkably, although superconductivity computed to disappear in the pristine bilayer reappears in the lithium intercalated bilayer with a significant boost. The Migdal-Eliashberg formalism leads to a $T_{\text c}$ above 11.3 K, which is significantly higher than the transition temperatures of monolayer and bilayer NiTe$_{2}$. The histograms of $\Delta_{\bk}$ for temperatures between 2 and 12 K are displayed in Fig.~\ref{fig3}(c), which again shows clearly two temperature-dependent gaps. The BCS fits for the two gaps are also plotted as solid lines in green and blue, respectively. Both of the gaps decrease with rising temperature, with $T_\text{c}$ of 11.1 and 11.3 K, respectively.  The zero-temperature superconducting gaps are, respectively, 3.1 and 1.8 meV, both significantly higher than the monolayer counterparts. The conspicuously boosted $T_{\text{c}}$ in LiNi$_{2}$Te$_{4}$ suggests that the intercalated Li plays decisive roles in the emerging superconductivity, which is to be clarified next.

 There is a significant electronic modification in the bilayer NiTe$_2$ intercalated with lithium. As seen in Fig. \ref{fig3}(d), LiNi$_{2}$Te$_{4}$ has a Fermi density of states  $N(0)= 4.7$ per eV (density of states is given on a per Ni basis), enjoying a near 2.5-fold increase over that of the pristine monolayer NiTe$_2$, where $N(0) = 1.9$ per eV). The intercalated lithiums donate nominally one electron per atom to the NiTe$_2$ bilayer. But doping alone is insufficient bring boost in the Fermi density of states. It can be seen from Fig. \ref{fig3}(d) that adding a half electron to the pristine monolayer increases $N(0)$ to a meager 2.4 per eV. The sharp peak at Fermi level in LiNi$_{2}$Te$_{4}$ is brought about by van Hove singularities along $\Gamma$-K path, as shown in Fig. \ref{fig3}(a), which results from lithium-mediated interlayer coupling. The increased density of states on the Fermi level means more electrons are susceptible to pairing interaction maybe mediated by dynamical phonons, which certainly contributes to the increased $T_{\text c}$.

\begin{figure}[h]
\centering
\includegraphics[width=76 mm]{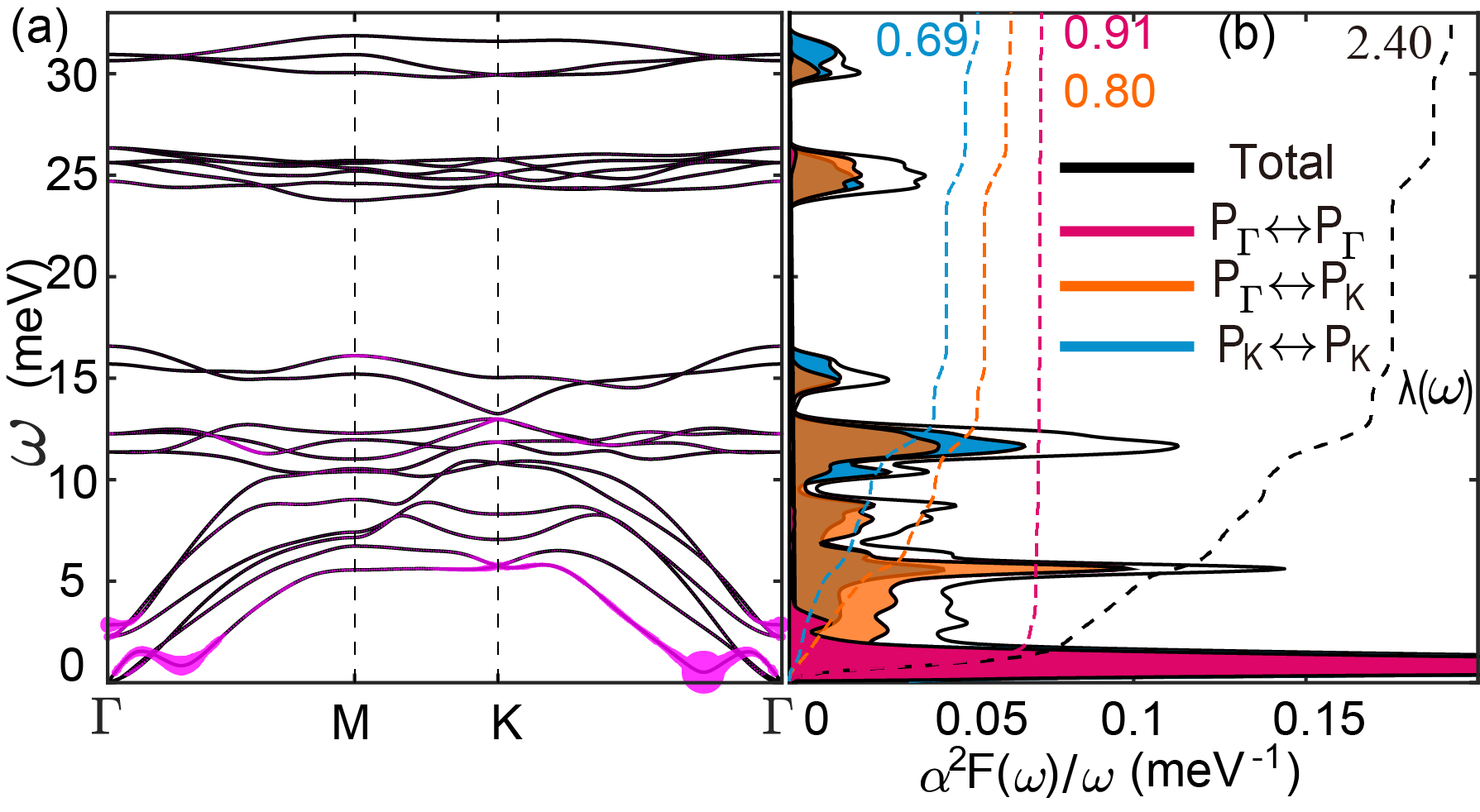} \\
\caption{(a) $\omega_{\mathbf{q}\nu}$ of LiNi$_{2}$Te$_{4}$, superimposed with $\lambda_{\mathbf{q}\nu}$, which is positively correlated to the size of the pink dot. (b) Total and partial  mass-enhancement parameter  $\gamma(P\leftrightarrow Q, \omega) = \alpha^{2}F(P \leftrightarrow Q, \omega)/\omega$, where ($P, Q=\Gamma, K$) (see main text). The corresponding accumulated electron-phonon coupling constants $\lambda(P \leftrightarrow Q, \omega)$ are depicted using colored dashed lines. }
\label{fig4}
\end{figure}

On top of multiplying the conducting electrons, the intercalated lithium also significantly enhances the electron-phonon coupling. This can be seen from the  overall mass-enhancement factor of LiNi$_{2}$Te$_{4}$, $\lambda=2.4$ as show in Fig.~\ref{fig4}(b), which is more than three-fold escalation from the pristine monolayer where $\lambda=0.74$. In the meantime, the presence of lithium, by virtue of its light mass and therefore large dynamical scale, gives rise to energetic phonons at around 31 meV.\citep{SM} This leads to an increase in $\omega_{\text{log}}$ by $\sim10$\%. Thus, even in the isotropic MAD estimate, it is expected an increase in the superconducting $T_{\text c}$. Indeed, the MAD approach gives a $T_{\text c} \sim 4.8 $~K,~\citep{SM} which is an inadequate estimate as the spectrum is evidently anisotropic.

A more detailed view of the electron-phonon coupling in the superconducting LiNi$_2$Te$_4$ can be gained by inspecting the $\bm k$-resolved mass-enhancement factor, 
\begin{equation}
	\lambda_{\bm k}=\frac{1}{N(0)} \int \frac{\mathrm{d} \omega}{\omega} \sum_{\bm k^{\prime}} \alpha^{2} F\left(\bm k, \bm k^{\prime}, \omega\right) \delta\left(\epsilon_{\bm k^{\prime}}\right),
\end{equation} 
where $\omega$ is phonon frequency, $\alpha^2$ the electron-phonon coupling strength averaged over the Fermi surface, and $F(\bm q, \omega)$ the phonon spectral function. The values of $\lambda_{\bm k}$ in the vicinity of Fermi surfaces are shown in Fig. \ref{fig1}(c) for monolayer NiTe$_2$ and in Fig. \ref{fig3}(b) for LiNi$_2$Te$_4$. This Migdal spectroscopy reveals a pervading enhancement of electron-phonon coupling in LiNi$_2$Te$_4$ in comparison with monolayer NiTe$_2$; that is, the dimensionless $\lambda_{\bm k}$ takes generally larger values in the former. In particular, the hole pockets centered at $\Gamma$ the electron-phonon coupling experience the strongest electron-phonon coupling, which arises from scattering with phonons in the low-energy regime, as been in Fig.~\ref{fig4}(b), where a sharp peak of partial mass-enhancement parameter  $\gamma(\Gamma\leftrightarrow\Gamma)$. Here, $\gamma(P\leftrightarrow Q)$ obtained by integrating $\alpha^2 F(\bm k, \bm k', \omega)/\omega$, restricting the momenta $\bk$ to the $P$ pocket and $\bk'$ to Q pocket [{\textit{cf.}} Eq. (1) and see Sec.~S2~\citep{SM}].  This softened optical modes along $\Gamma$-K path indicate an impending charge-density wave, which makes the low-frequency phonons  scatter strongly the holes near $\Gamma$, leading to large $\lambda_{\bm k}$ in this region.

In summary, using {\it ab initio} calculation based on anisotropic Migdal-Eliashberg theory within the static lattice approximation, we have shown the emergence of two-gap superconductivity in two-dimensional NiTe$_{2}$ crystals, namely, monolayer NiTe$_{2}$ and lithium intercalated bilayer, although the bulk NiTe$_{2}$ does not superconduct. The energetically most favored lithium intercalated bilayer, LiNi$_{2}$Te$_{4}$, is computed to have a relatively high superconducting transition $T_{\text c}$  up to 11.3~K with superconducting gaps up to 3.1 meV, whereas the monolayer has a $T_{\text c}$ of 5.7~K and the pristine bilayer NiTe$_{2}$ is nonsuperconducting. The $T_{\text c}$ of the Li interacalted bilayer is highest among all known phonon-mediated two-dimensional superconductors exfoliated from layered materials,\citep{Ugeda15,Navarro-Moratalla16} and among layered materials, it is only second to MgB$_2$~\citep{Choi02} and on a par with Ca intercalated graphite.\citep{Weller05} The remarkable enhancement of superconductivity in the lithium intercalated bilayer  is attributable to the synergy of electron doping and lithium-mediated interlayer hybridization and the accompanying electron-phonon coupling, as well as energetic phonons due to the intercalated lithium. It should be remarked that given the exceedingly strong coupling between the lattice and electron in the Li intercalated bilayer, the normal state may become a non-Fermi liquid, which is not captured in the current theoretical framework but will be a highly intriguing scenario for further investigation.

 Finally, remarks on the experimental preparation of these emergent highly crystalline two-dimensional superconductors are in order. The NiTe$_2$ monolayer and bilayer have already been experimentally synthesized,\citep{Zhao18} and their superconductivity and the absence thereof can be straightforwardly probed. The lithium intercalation to the bilayer can be achieved with ionic liquid gating.~\citep{Klipstein87,Yu15,Sajadi18,Fatemi18} Therefore, the experimental accessibility, combined with relatively high $T_\text{c}$, substrate independence and proximity tunability will make these superconductors ideal platforms for exploring intriguing correlation effects and quantum criticality associated two-dimensional superconductivity. We have also examined intercalation of bilayer NiTe$_2$ with sodium and potassium, with corresponding $T_{\text c} =$ 4.4 and 3.1~K, respectively,\citep{SM} which are also interesting possibilities worth experimental assaying.

\begin{acknowledgements} 
This work is supported by National Natural Science Foundation of China (11804118, 11725415, 11934001, 21703081 and 21973034), Ministry of Science and Technology of the People's Republic of China (2018YFA0305601, 2016YFA0301004), Strategic Priority Research Program of Chinese Academy of Sciences, Grant No. XDB28000000, the Fundamental Research Funds for the Central Universities (21617330), and by the Natural Science Foundation of Guangdong Province (2018A030313386).  The Calculations were performed on the Tianhe-I Supercomputer System and high-performance computation cluster of Jinan University.

F.P. Zheng and X.-B. Li contribute equally to this work.
\end{acknowledgements}


\end{document}